\documentclass[twocolumn,superscriptaddress,floatfix,preprintnumbers,prl ,nofootinbib,hyperref,longbibliography]{revtex4-1}
\usepackage{enumerate}
\usepackage{mathrsfs,amsmath,amssymb}
\usepackage{natbib}
\usepackage{graphicx}
\usepackage{cancel}
\usepackage[normalem]{ulem}
\usepackage[colorlinks, linkcolor=red, anchorcolor=blue, citecolor=blue, colorlinks=true, urlcolor=blue]{hyperref}
\hypersetup{colorlinks=true}
\usepackage{stackrel}
\usepackage{pgfplots}
\usepackage{subfigure}
\usepackage{makecell}
\usetikzlibrary{positioning}
\usetikzlibrary{snakes}
\usepackage{multirow}
\usepackage{slashed}
\usepackage{float}
\newcommand{\prlsection}[2]{{\it\textbf{#1}{#2}}---}

\newcommand\be{\begin{equation}}
\newcommand\ee{\end{equation}}
\newcommand\bea{\begin{eqnarray}}
\newcommand\eea{\end{eqnarray}}

\begin{document}

\title{ Chiral Magnetic Effect induced Spectator Process for Leptogenesis}
\author{Wei Chao}
\email{chaowei@bnu.edu.cn}
\affiliation{Key Laboratory of Multi-scale Spin Physics, Ministry of Education, Beijing Normal University, Beijing 100875, China}
\affiliation{Center of Advanced Quantum Studies, School of Physics and Astronomy, Beijing Normal University, Beijing, 100875, China }
\vspace{3cm}

\begin{abstract}

Conventional Leptogenesis mechanism, which provides compelling explanation to the origin of the baryon asymmetry of the universe (BAU), assumes the absence of hypermagnetic field in the early universe, thereby disregard the implications of hyper gauge field helicity, that have been thoroughly studied in the magnetogenesis mechanism.  
In this paper, we address impacts of a general U(1) gauge field on Leptogenesis by deriving equation of motions for the helicity and the energy density of a general magnetic field, to which the chiral magnetic effect (CME) is identified as essential, and studying their effects on the evolution chiral asymmetries. 
Notably, CME in the $U(1)_{\mathbf{L}_i-\mathbf{L}_j}$ framework, where $\mathbf{L}_{i,j}$ means specific lepton flavor, explicitly breaks the total lepton number and provides an efficient spectator process, that can wash out  pre-existing lepton asymmetries. 
This establishes a natural connection to the wash-in Leptogenesis paradigm. 
We demonstrate that this spectator effect enables the generation of the BAU, eliminating the need for both an initial $\mathbf{B}-\mathbf{L}$ charge and primordial helicity.

\end{abstract}

\maketitle

\prlsection{Introduction}{.}  The baryon asymmetry of the universe (BAU) stands as a key evidence for new physics beyond the Standard Model (SM).  The BAU,  measured at the time of the Big Bang Nucleosynthesis and the Cosmic Microwave Background~\cite{Planck:2018vyg}, is 
\bea
\eta_B = \frac{n_b }{s} = 9 \times 10^{-11} \; ,
\eea
where $n_b$ and $s$ are the baryon number density and the entropy density, respectively. 
To address this value, a dynamical theory must satisfy three Sakharov conditions~\cite{Sakharov:1967dj}: Baryon number violation, C and CP violation and a departure from the thermal equilibrium. However, the second and the third conditions are not satisfied in the minimal SM, which catalyzes the model buildings for Baryogenesis.

Many workable {\it Baryogenesis} mechanisms~\cite{Fukugita:1986hr,Cohen:1993nk,Trodden:1998ym,Morrissey:2012db,Affleck:1984fy,Cohen:1988kt,Chao:2017oux,Co:2019wyp,Domcke:2020kcp,Domcke:2019mnd,Croon:2019ugf,Davoudiasl:2010am,Shelton:2010ta,Cui:2011ab,Hall:2010jx,Elor:2018twp,Chao:2023ojl,Elor:2022hpa,Domcke:2020quw,Marshak:1979fm,Mukaida:2021sgv,Barrie:2021mwi,Cui:2021iie,Dror:2019syi} have been proposed, of which the Leptogenesis mechanism has received widespread attention due to the flourishing development of the neutrino physics. 
In Leptogenesis, the CP-violating decay of the lightest seesaw particle  gives rise to the production of a net $\mathbf{B-L}$ charge, which is partially washed-out by processes mediated by seesaw particles and is finally converted into the baryon asymmetry  and the lepton asymmetry via the weak sphaleron process, which violates $\mathbf{B}$ and $\mathbf{L}$ separately, but keeps $\mathbf{B-L}$ conservation.    
Recently, some fantastic achievements have been obtained and many new models relevant to Leptogenesis have been proposed, such as the gravitational wave Leptogenesis~\cite{Alexander:2004us,Lyth:2005jf,Lambiase:2006md,Maleknejad:2016dci,Adshead:2017znw,Kamada:2019ewe}, axion-inflation Leptogenesis~\cite{Anber:2015yca,Adshead:2016iae,Jimenez:2017cdr,Domcke:2019mnd,Domcke:2022kfs,Chao:2024fip,Fukuda:2024pkh},  wash-in Leptogenesis~\cite{Domcke:2020quw}, Affleck-Dine Leptogenesis~\cite{Barrie:2021mwi,Barrie:2022cub}, Eogenesis~\cite{Chao:2024uxa}, Lepto-axiogenesis~\cite{Co:2019wyp,Co:2020jtv,Kawamura:2021xpu,Barnes:2022ren,Barnes:2024jap,Chao:2023ojl} etc.  These new Leptogenesis mechanisms either modify the spectator processes for Leptogenesis or provide new CP-violating source terms.  From this point of view, Leptogenesis is now in an era of flourishing diversity.

In this paper, we focus on the impact of a general U(1) gauge interaction on the Leptogenesis mechanism. The change of the total $\mathbf{B}$ and $\mathbf{L}$ over a finite time interval  in the minimal SM is given by~\cite{Adler:1969gk,Bell:1969ts,tHooft:1974kcl}
\bea
\Delta \mathbf{B} =\Delta \mathbf{L} = N_f \Delta N_{CS} -N_f \frac{g^{\prime 2}}{16\pi^2} \Delta {\cal H}_Y
\eea 
where $N_f$ is the generation of the SM fermion, $N_{\rm CS}^{}$ is the Chern-Simons number of $SU(2)_L$ and ${\cal H}_Y$ is the helicity of the hypermagnetic field.  It was shown that~\cite{Fujita:2016igl}  non-zero ${\cal H}_Y$  can produce the observed amount of baryon asymmetry through the chiral anomaly without any ingredients beyond the SM. In the meanwhile the chiral magnetic effect (CME)~\cite{Vilenkin:1980fu,Fukushima:2008xe,Kharzeev:2013ffa}, that is  an electric current proportional to the chiral chemical potential can be induced when a system with chiral asymmetries is subjected to a magnetic field, can lead to the suppression of the generation of the BAU in cases with large magnetic field strength~\cite{Kamada:2016eeb}.   Considering that anomalies induced by the  $U(1)_Y$  keeps $\mathbf{B-L}$ conservation,  interplaies between the hypermagnetic field and the BAU have turned out to involve a qualitative uncertainty~\cite{Hamada:2025cwu},  not even to say the effect of  non-helical hypermagnetic field.   We investigate BAU in the framework of SM extended by the gauged $U(1)_{\mathbf{L}_i-\mathbf{L}_j}$~\cite{He:1991qd}, where $\mathbf{L}_{i,j}$ stands for specific family of leptons. Notably, chiral anomalies induced this U(1) gauge interactions violate $\mathbf{B-L}$ explicitly. As a result, the helical magnetic field of the $U(1)_{\mathbf{L}_i-\mathbf{L}_j}$ can lead to the production of the net $\mathbf{B-L}$ charge, while the non-helical magnetic field of the $U(1)_{\mathbf{L}_i-\mathbf{L}_j}$ can lead to a new spectator process for Leptogenesis via the CME, which has the same effect as the wash-out process in the traditional Leptogenesis mechanism. 
We derive transport equations for various SM fermions, the helicity and the energy density of new magnetic field, and carry out full numerical calculations, which show that  the BAU can be addressed by this new gauge interaction in the presence of either helical magnetic field or non-helical magnetic field with initial chiral asymmetries that  keep  $\mathbf{B}-\mathbf{L}=0$.

The remaining of the paper is organized as follows: in the section II  we study chiral anomalies of the $U(1)_{\mathbf{L}_i-\mathbf{L}_j}$.  We derive transport equations for SM fermion, the helicity and the energy density of the new magnetic field in section III, then we calculate the BAU by solving transport equations numerically. The last part is concluding remarks.

\prlsection{$U(1)_{\mathbf{L}_e-\mathbf{L}_\mu}^{}$ and Triangle anomalies}{.} \label{sec:nonrestoration} In this section we  give a brief overview on the $U(1)_{\mathbf{L}_e-\mathbf{L}_\mu}^{}$ gauge symmetry  and calculate triangle anomalies associated to it.
U(1) extensions to the SM are a fertile area of research, combining theoretical elegance with testable predictions for collider, precision, and cosmological experiments.
Gauged $U(1)$ models~\cite{Langacker:2008yv} usually require to introduce extra fermions for anomaly cancellations. The gauge symmetries related to baryon number or lepton number, such as $\mathbf{B}$, $\mathbf{L}$~\cite{FileviezPerez:2010gw,Dulaney:2010dj,Chao:2010mp}, $\mathbf{B+L}$~\cite{Chao:2015nsm,Chao:2016avy}, require vector-like fermions as well as right-handed neutrinos  for anomaly cancellation,  The $U(1)_\mathbf{B-L}$~\cite{Mohapatra:1980qe,Marshak:1979fm,Wetterich:1981bx} and $U(1)_\mathbf{R}$~\cite{Chao:2017rwv}, where the subscript R means right-handed fermions, only need three right-handed neutrinos for anomaly cancellation,  while $U(1)_\mathbf{L_i-L_j}$~\cite{He:1991qd}  does not need any extra fermions for the anomaly cancellation.  
In this work we focus on the $U(1)_{\mathbf{L}_e-\mathbf{L}_\mu}^{}$ model, in which charges of various fermions are listed in the table.~\ref{tab:merged_cells}.   It is easy to check that all anomalies, i.e., ${\cal A}_1 (SU(3)_C^2 \otimes U(1)_X)$, ${\cal A}_2 (SU(2)_L^2 \otimes U(1)_X)$, ${\cal A}_3 (U(1)_Y^2 \otimes U(1)_X)$, ${\cal A}_4 (U(1)_Y \otimes U(1)^2_X)$, ${\cal A}_5 ( U(1)_X^3)$, ${\cal A}_6 (U(1)_X)$~\cite{Witten:1982fp,Adler:1969gk,Bell:1969ts,Bardeen:1969md,Eguchi:1976db,Alvarez-Gaume:1983ihn}, are automatically cancelled in the minimal SM.

\begin{table}[t]
    \centering
    \begin{tabular}{c|c|c|c|c|c}
        \hline
        \hline
        \multirow{1}{*}{Fermions} & $L_e$ & $L_\mu$ & $E_e$ & $E_\mu $ & Others \\
        \hline
     $U(1)_{\mathbf{L}_e-\mathbf{L}_\mu}$ charge & 1 & 1 & -1 & -1 & 0 \\
        \hline
        \hline
    \end{tabular}
     \caption{Charge settings of various SM fermions in the $U(1)_{L_e -L_\mu}$ model, where $L_{e/\mu}$ and $E_{e/\mu}$ stands for left-handed and right-handed fermions of the first two generations, respectively. }
    \label{tab:merged_cells}
\end{table}

To estimate its impact to primordial physics, we consider triangle anomalies of  chiral fermions in  $U(1)_{\mathbf{L}_e-\mathbf{L}_\mu}$.  Following the standard strategy,  triangle anomalies for  the first two generation leptons can be calculated as 
\bea
\partial_\mu  j^\mu_{L_{e/\mu}} &=& \frac{1}{32 \pi^2 } \left(  g^2 W \widetilde W +g^{\prime 2} F \widetilde{F} +4  g^{ 2 }_{X} X \widetilde{X}\right)\\
\partial_\mu j^\mu_{E_{e/\mu}} &=& \frac{1}{16 \pi^2 } \left( - g^{\prime 2} F \widetilde{F} - g^{ 2 }_{X} X \widetilde{X}\right) 
\eea
where $W\widetilde{W}$, $F\widetilde{F}$ and $X \widetilde{X}$ are Chern-Simons (CS) term for the $SU(2)_L$ gauge field,  the $U(1)_Y$ gauge field and the the $U(1)_{\mathbf{L_e-L_\mu }}$ gauge field, respectively. 
Other triangle anomalies are listed in the Appendix A.
Following the definition of the baryon current and the lepton current as  $j_B^\mu = {1\over 3} \sum_f \left( \bar Q \gamma^\mu Q + \bar u_R \gamma^\mu u_R^{}\right.$ + $\left.\bar d_R^{} \gamma^\mu d_R^{}  \right)$ and $J_L^\mu = \sum_f \left( \bar \ell_L \gamma^\mu \ell_L^{} + \bar E_R^{} \gamma^\mu E_R^{} \right)$, one has 
\bea
\partial_\mu j^\mu_{\mathbf{B}-\mathbf{L}}  = -{g_X^2 \over 8 \pi^2  } X \widetilde{X} \; , \label{master0}
\eea
which is the key formula for the baryogenesis.   We will present two possible ways of generating the BAU via this term in the next section.  This mechanism shares merits of the gravitational wave Leptogenesis mechanism, which takes the CS term of gravitational waveas the source term, that does not work in the case neutrino being Dirac particles.

\prlsection{BAU}{.} \label{sec:bau}    In this section we study the BAU in the presence of the $U(1)_{\mathbf{L}_e-\mathbf{L}_\mu}$ gauge symmetry. Before moving to specific model, lets first focus on transport equations of chemical potential for the first two generation leptons. The equation of motion for comoving chemical potentials, which is proportional to the number density of particle minus anti-particle divided by the entropy density, are given by
\bea
-\frac{d}{d \ln T} \left( \frac{\mu_{E_i}}{T} \right) &=& -{1\over g_{E_i}} {\gamma_{E_i} \over H} \left( \frac{\mu_{E_i}}{T} -\frac{\mu_{L_i}}{T} + \frac{\mu_H}{T} \right)  \nonumber \\
&& + q_e^2 \gamma_y + q_l^2 \gamma_l  \label{master1-0}\\
-\frac{d}{d \ln T} \left( \frac{\mu_{L_i}}{T} \right) &=& +{1\over g_{E_i}} {\gamma_{E_i} \over H} \left( \frac{\mu_{E_i}}{T} -\frac{\mu_{L_i}}{T} + \frac{\mu_H}{T} \right)  \nonumber \\
&&-\frac{1}{g_{L_i}} \frac{ \gamma_{WS}}{H} \left[ \sum_i {\mu_{L_i} \over T} + 3 \sum_i {\mu_{Q_i} \over T} \right] \nonumber \\
&& - q_L^2 \gamma_y - q_l^2 \gamma_l \label{master1-2}
\eea 
where $H$ is the Hubble rate, $\gamma_{E_i}$ is the rate of lepton Yukawa interaction, $\gamma_{WS}$ is the weak sphaleron rate, $\gamma_y$ is the term from the helical hyper-magnetic field and $\gamma_l $ is the term from  helical magnetic field of new U(1) gauge symmetry.  Transport equations for the third generation lepton as well as various quarks are the same as these in the SM.  It was claimed that primordial helical hyper-magnetic field can solely generate large enough BAU. Considering that this term does not contribute to the total $\mathbf{B}-\mathbf{L}$, this mechanism may only work in certain  extreme condition. In this work, we ignore the effect of the $U(1)_Y$ on the BAU,  by assuming that the primordial hyper-magnetic field is negligible, and only consider the contribution of the $U(1)_{\mathbf{L}_e -{L}_\mu}$.  


\begin{figure*}[t]
        \includegraphics[width=8cm]{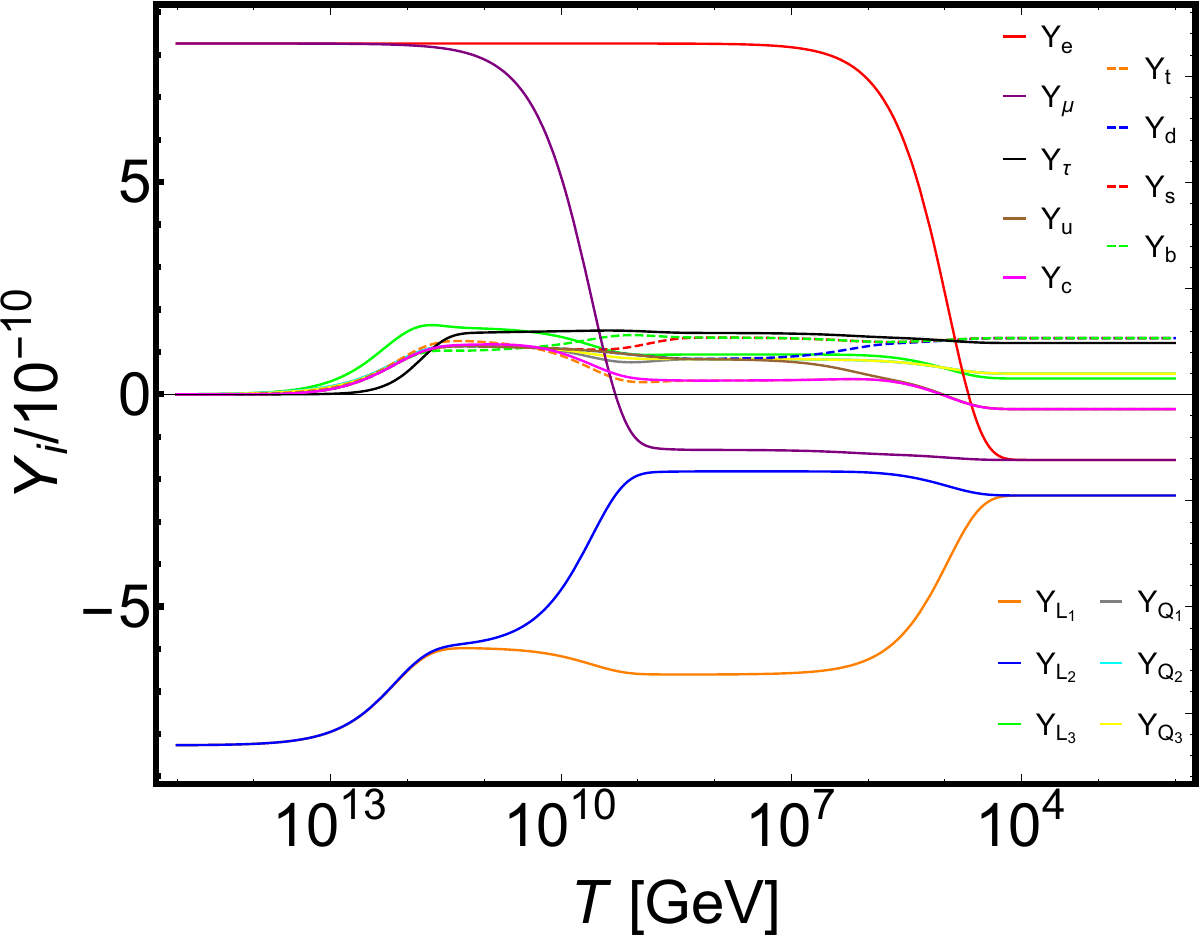}
         \includegraphics[width=8cm]{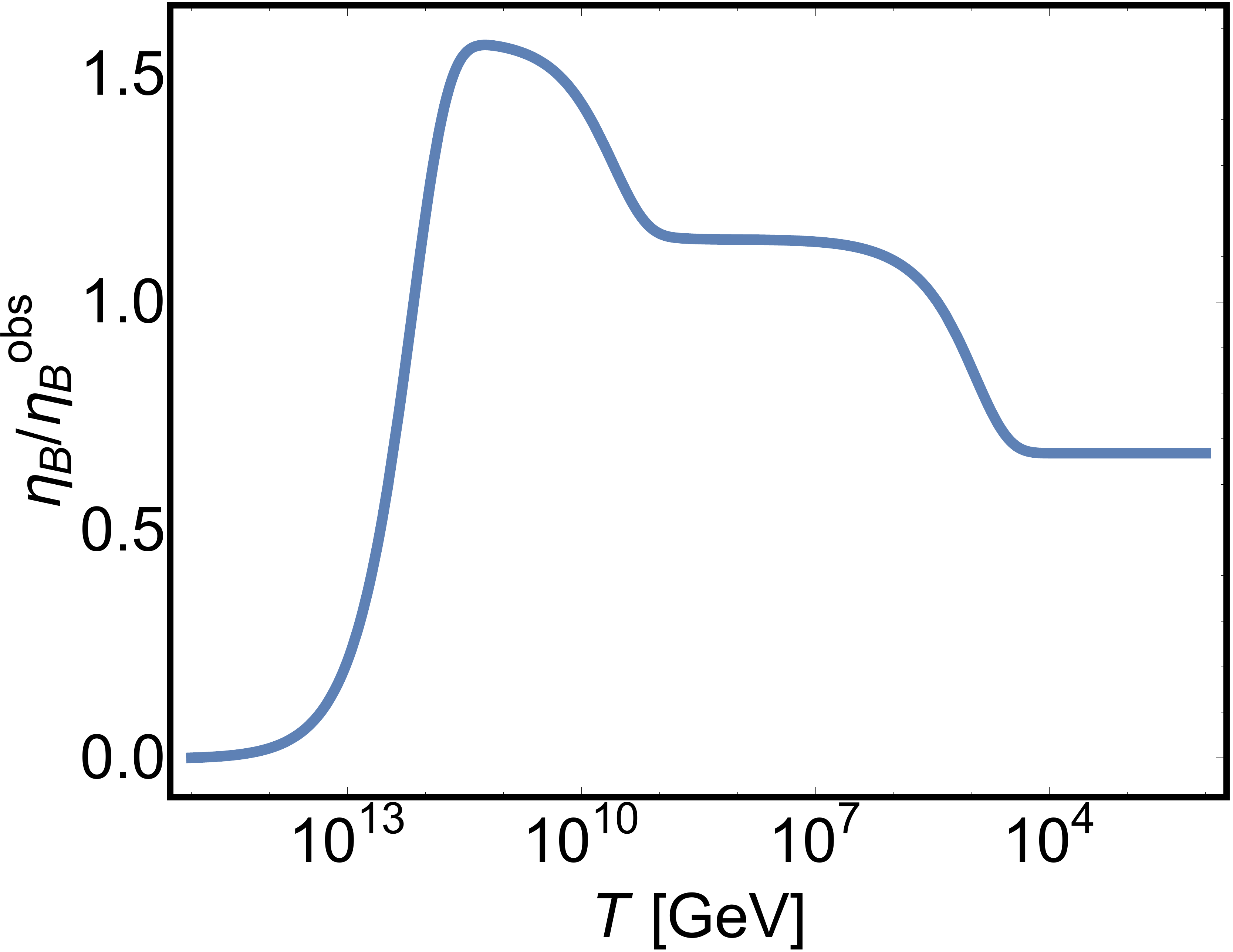}
	\caption{ Left-panel: $Y_i=\mu_i/T$ as the function of the temperature in the presence of helical magnetic field; Right-panel: $\eta_B/\eta_B^{\rm obs}$ as the function of the temperature.}
	\label{Afig}
\end{figure*}

The time derivative  of the  $U(1)_{\mathbf{L}_e -\mathbf{L})\mu}$ magnetic helicity is only term contributing to transport equations, considering that the vacuum structure of an Ablein gauge field is trivial, and  $\gamma_l$ can be written as 
\bea
\gamma_i^{}=- \frac{6 }{g_i H T^3}  \frac{\alpha^\prime}{2\pi} \dot{h}_p  \label{master2}
\eea
where  $h_p \equiv \lim_{V \to \infty} {{\cal H } \over V }$ being the new magnetic helicity density with ${\cal H}$~\cite{moffatt1978} the new magnetic helicity, whose expression is given in the appendix. Using Ampere’s law and the generalized Ohm’s law~\cite{Figueroa:2017hun}, the time evolution of the conformal helicity density $h_c$, which is related to $h_p$ via $h_c = a^3 h_p$~\cite{Jimenez:2017cdr} with $a$ the scale factor, can be written as~\cite{Domcke:2019mnd}
\bea
\frac{d h_c}{d\eta } = \lim_{V\to \infty} \int{d^3 x \over V} \left( \frac{2}{\sigma}B \cdot \nabla^2 A + \frac{4\alpha^\prime}{\pi} \frac{\mu_X}{\sigma} B^2 \right) \label{master3}
\eea
where $A$  and $B$ are the vector potential and the magnetic field strength, respectively. $\sigma$ is the conductivity~\cite{Arnold:2000dr} of the plasma in the conformal frame,  $\sigma = \sigma_p a$.  Following the strategy of the Ref.~\cite{Laine:2005bt,RostamZadeh:2015xnd,Rogachevskii:2017uyc}, the coefficient $\mu_X$ can be written as
\bea
\frac{\mu_X}{a} = -2 \mu_{L_1}^{} -2 \mu_{L_2}^{} + \mu_{e_1}^{}+ \mu_{e_2}^{} \; .
\eea
%
Combining Eqs(\ref{master1-0}), (\ref{master1-2}) with Eqs.(\ref{master2}) and (\ref{master3}), one can estimate impact of  the helical or non-helical new magnetic field to the BAU.  Generally speaking, two effects can be induced: (1) The new CS term provides the initial source terms for the first two generation leptons  via the  so-called axion baryogenesis mechanism; (2) The new CS term provides a new spectator process in the presence of external non-helical magnetic field.  We will study these two effects in the following.

\begin{figure*}[t]
        \includegraphics[width=8.5cm]{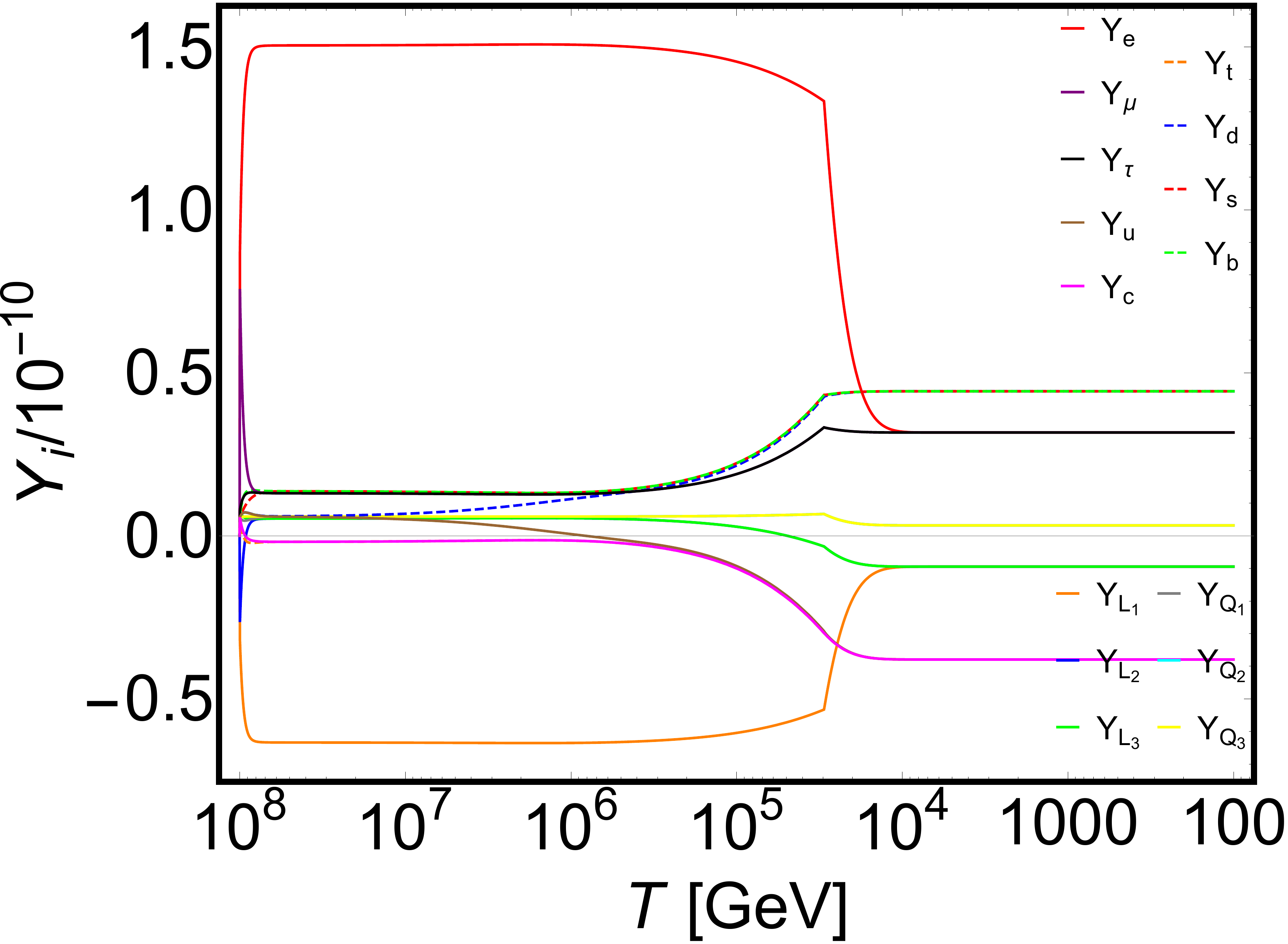}
        \includegraphics[width=8.5cm]{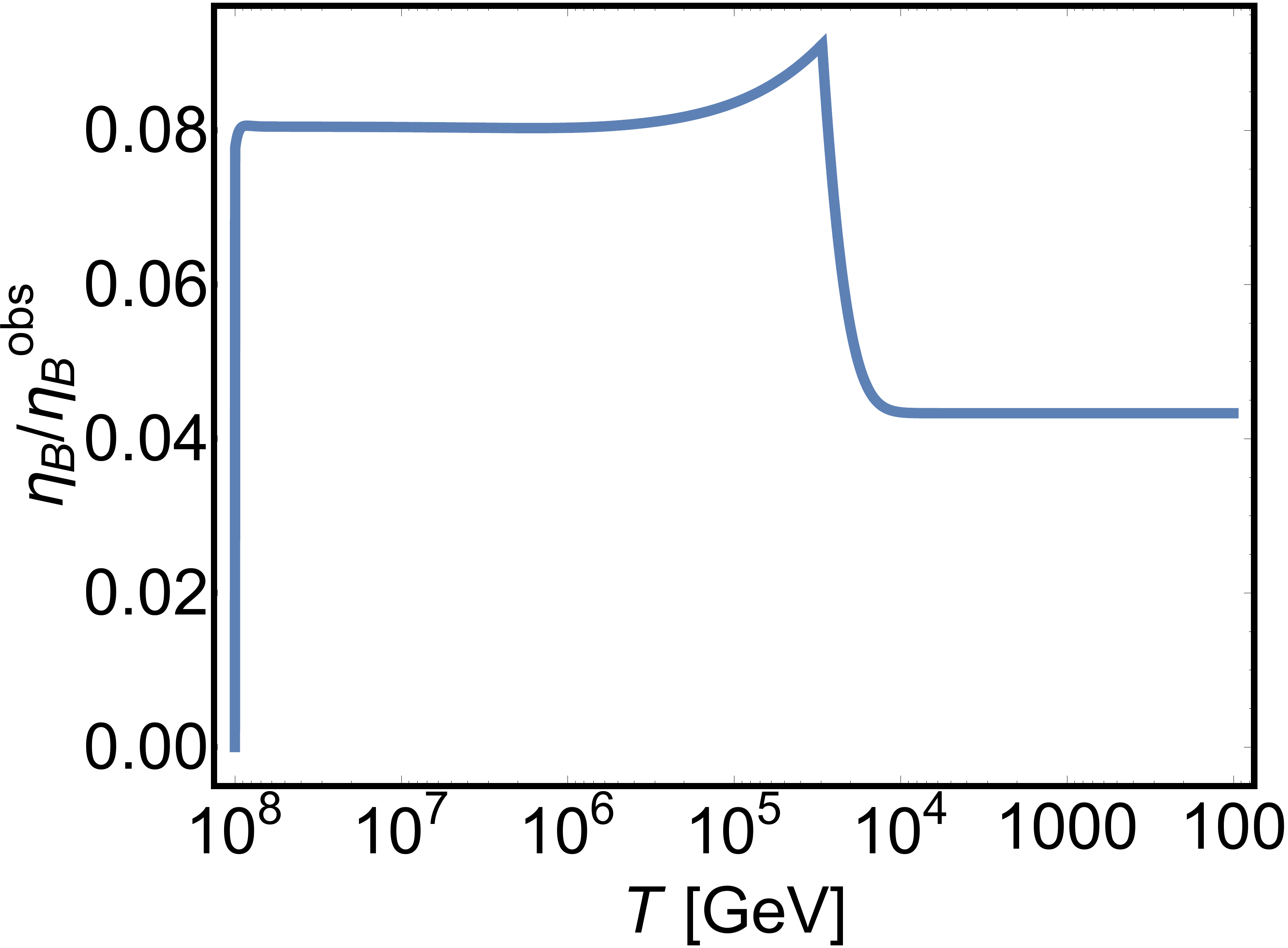}
	\caption{ Left-panel: $Y_i=\mu_i/T$ as the function of the temperature, with the meaning of each curve explained in the main text; Right-panel: $\eta_B /\eta_B^{\rm obs}$ as the function of the temperature. }
	\label{fig2-A}
\end{figure*}

\noindent \framebox[\width]{\textbf{\color{red}~CS term as the source term:~}} The axion-inflation Baryogenesis mechanism~\cite{Anber:2015yca,Adshead:2016iae,Jimenez:2017cdr,Domcke:2019mnd,Domcke:2022kfs,Chao:2024fip,Fukuda:2024pkh} provides  a theoretical framework that unifies three key cosmological phenomena: cosmic inflation, axion physics, and baryogenesis. 
The basic idea is that the axion-like inflaton, that couples to a CS term of gauge fields or graviton, generates the helical magnetic field as well as the CS number during the inflation. 
The CS number can be converted into the initial number densities of chiral plasmas via the triangle anomaly during the reheating epoch or during the subsequent evolution of the helical magnetic field in the radiation dominated universe. 
These non-zero initial number densities of chiral plasmas will finally be converted into non-zero BAU via the electroweak sphaleron process, which breaks the $\mathbf{B}$ and $\mathbf{L}$, but keeps $\mathbf{B}-\mathbf{L}$ conservation.  
It has been shown that The observed BAU can be addressed if large enough non-vanishing helical magnetic field stored in the  hyper-magnetic field survives until the electroweak phase transition (EWPT) and sources the BAU after the quench of the electroweak sphaleron process.  
%
The initial number densities of plasmas produced from the CS number of hyper-magnetic field keep B-L conservation.
On the other hand, if the axion inflaton couples to $X\widetilde{X}$,  the new CS number produced during the inflation does not keep $\mathbf{B-L}$ conservation at all, as can be seen from the eq. (\ref{master0}),  and  it naturally generates an initial B-L asymmetry, which gives rise to the  BAU.

We show in the left-panel of the Fig.~\ref{Afig} various conformal chemical potentials  as the function of the temperature, where the orange, blue, green, red, purple, black, gray, cyan, yellow, brown, magenta, dashed-orange, dashed-blue, dashed-red, dashed-green curves corresponds the running behavior of  left-handed lepton doublets, right-handed lepton singlets, left-handed quark doublets  and right-handed up-type quarks and down-type quarks, respectively. We have assumed that initial chiral densities for  the first two generation leptons are generated during the reheating temperature by the CS number generated during the axion inflation, with $Y_{e_{1/2}, L_{1/2}}=\pm 8.3\times 10^{-10}$.  Typical temperatures correspond to  changes of the shape of each curve are consistent with the analytical estimation of characteristic  equilibrium temperature. Obviously, non-zero BAU can be generated in this case,  with numerical results shown right-panel of the Fig.~\ref{Afig}.

\noindent \framebox[\width]{\textbf{\color{red}~CS term triggered spectator process:~}} There are two important ingredients in the Leptogenesis: the CP-violating  (CPV) source term and the spectator process~\cite{Davidson:2008bu}. 
In the traditional Leptogenesis mechanism the out of equilibrium decay of right-handed neutrinos generates non-zero B-L number density. It is then converted into the BAU via the electroweak sphaleron process which is taken as the spectator process. 
Recently, it has been shown that the process mediated by the heavy neutrinos can be taken as a new spectator process and  address the BAU via the so-called wash-in Leptogenesis mechanism~\cite{Domcke:2020quw}.  
Actually, if looking into the eq.~(\ref{master3}), one can find that the first term on the right-handed side may provide a source term for BAU in the presence of helical magnetic field as can be seen from previous discussions, while the second term may provide a new spectator process for Leptogenesis in the presence of non-helical magnetic field. 

Let us check this point by performing full numerical calculations of transport equations in the presence of non-helical magnetic field and non-zero initial chiral densities that keep $\mathbf{B}-\mathbf{L} =0$.  It was well-known that null initial  $\mathbf{B}-\mathbf{L}$ density results in null BAU if the electroweak sphaleron process is the only spectator process, otherwise new spectator process is at work.  
We assume that the scale for spontaneous breaking of the $U(1)_{\mathbf{L}_e-\mathbf{L}_\mu}$ symmetry is lower than the electroweak scale, resulting in a low mass gauge field, such that the new gauge field may have non-trivial affect on the evolution of the BAU.
It is convenient to perform a fourier transformation to the helicity density and the energy density of the magnetic field, the detail of which is given in the Appendix B.  Then transport equations for the $k$-mode of helicity density and the energy density are 
\bea
-\frac{\partial h_k}{\partial \ln T } &=& -\frac{2k^2}{\sigma_0} \frac{T}{H} h_k + \frac{8 \alpha^\prime}{\pi\sigma_0} \frac{T}{H}\frac{\mu_{X}^{}}{T} \rho_k \label{master4} \\
-\frac{\partial \rho_k}{\partial \ln T} &=& -\frac{2k^2}{\sigma_0}\frac{T}{H} \rho_k + \frac{2 \alpha^\prime k^2 }{\pi\sigma_0} \frac{T}{H} \frac{\mu_{X}^{}}{T} h_k 
\eea
where the first term is due to the diffusion leading to the decay of magnetic helicity, the second term is responsible for the chiral plasma instability and can be taken as a new spectator term. It is obvious that this process does not work in cases either initial chiral densities or initial magnetic field is zero, which is  is what distinguishes it from the conventional spectator process.  Following Refs.~\cite{Baym:1997gq,Arnold:2000dr}, we take the electric conductivity as $\sigma =a T /\alpha^\prime$ in the numerical analysis.

Notice that a high wave number gives  significant  impact on the chiral plasma stability. However magnetic field does not survive the Ohmic dissipation plasma for a much high wave number. Here we follow the discussion of Ref.~\cite{Elahi:2021pug}, and set $k_{\rm max}^{} \sim 10^{-7} T_{\rm EW}^{} $ in our numerical analysis.  
We show in the left-panel of the Fig.~\ref{fig2-A},  various conformal chemical potentials as the function of the temperature  by setting initial  number densities for leptons as $ n_{L_1}^{}= 1/2 n_{R_1}^{}=0.51\times 10^{-10}$.  
The total B-L is exactly zero in this case, and the non-zero B-L charge can not be produced in the conventional Leptogenesis case.  
In our case, due to the existence of non-zero initial chiral asymmetry and the non-helical magnetic field, the new spectator process given in the second term on the right-handed side of the Eq. (\ref{master4})  that violates B-L comes into play a rule.  The charge density of various chiral fermions is correspondingly changed and the non-zero BAU is generated, as shown in the plot. The relic of the helicity depends on the initial input of  the non-helical magnetic field and the chiral asymmetries of plasma.  The evolution of the BAU is illustrated in the right-panel of the Fig.~\ref{fig2-A}, and non-zero BAU is generated in this scenaro.

\prlsection{Summary}{.} Leptogenesis mechanism has made significant progress in various aspects since it was first proposed in 1986. In this paper, we pointed out the potential contribution to the Leptogenesis induced by the CS term of the $U(1)_{\mathbf{L}_e-\mathbf{L}_\mu}$ gauge symmetry.  Considering that the CS term relevant to $U(1)_{\mathbf{L}_e-\mathbf{L}_\mu}$ violates $\mathbf{B}-\mathbf{L}$ explicitly,  its effect on baryogenesis is unique when compared with other U(1) gauge symmetries. We have shown the following two scenarios that can give rise to successful baryogenesis:
\begin{itemize}
\item  Helical magnetic field is produced via the axion-inflation and the $U(1)_{\mathbf{L}_e-\mathbf{L}_\mu}$ is spontaneous broken during the reheating epoch. In this case, chiral fermions relevant to the first two generation leptons  are produced from the decay of  magnetic helicity, which is then transported into the baryon asymmetry via the electroweak sphaleron  process. 
\item   Non-helical magnetic field is present in the early universe and the $U(1)_{\mathbf{L}_e-\mathbf{L}_\mu}$ is spontaneous broken after the electroweak phase transition. We found that the CME effect will  lead to a lepton-number-violating spectator process for  leptogensis, which is similar to wash-out processes mediated by right-handed neutrinos in the wash-in Leptogenesis mechanism. In this case a non-zero B-L charge can be obtained, even though  the initial B-L charge is zero. 
\end{itemize}    
These two results make the U(1) extensions of the SM a possible way to address the BAU. It should be mentioned that other U(1)  gauge symmetries, such as  can  B, L, B-L etc., can have similar effect on baryogenesis for the case neutrino being Dirac particles, in which transport equations of right-handed neutrinos are decoupled  from those of SM particles. 
\label{sec:summ}

\prlsection{Acknowledgments}{.} This work was supported by the National Natural Science Foundation of China (NSFC) (Grants No. 12447105, No. 11775025 and No. 12175027). and the Fundamental Research Funds for the Central Universities under grant No. 2017NT17.

\section{appendix A} 
Current equations for various quarks and the third generation lepton takes the following form
\bea
\partial_\mu \left( j^\mu_{B, Q} \right)&=& \frac{1}{32 \pi^2 } \left(  g^2 W \widetilde W + \frac{1}{9}g^{\prime 2} F \widetilde{F} \right)\\
\partial_\mu \left( j^\mu_{B, u} \right)&=& \frac{1}{16 \pi^2 } \left(  - \frac{4}{9}g^{\prime 2} F \widetilde{F} \right) \\
\partial_\mu \left( j^\mu_{B,d} \right)&=& \frac{1}{16 \pi^2 } \left(   -\frac{1}{9}g^{\prime 2} F \widetilde{F}\right) \\
\partial_\mu \left( j^\mu_{L_\tau~~} \right)&=& \frac{1}{32 \pi^2 } \left(  g^2 W \widetilde W +g^{\prime 2} F \widetilde{F} \right)\\
\partial_\mu \left( j^\mu_{E_{\tau~~}} \right)&=& \frac{1}{16 \pi^2 } \left( - g^{\prime 2} F \widetilde{F} \right) \; , 
\eea
which are the same as those in the SM. 

\section{appendix B}

In this section we derive equations that govern the evolution of the helicity.  The helicity density is defined by 
\bea
h(t) = \lim_{V\to \infty} \frac{1}{V} \int d^3 x \mathbf{A} \cdot \mathbf{B}
\eea 
where $\mathbf{A}$ is the vector potential, $\mathbf{B}$ is the magnetic field strength.  
Defining the vector potential as
\bea
A=\int \frac{d^3 k }{(2\pi)^3} \sum_\lambda \varepsilon_k^\lambda A_k^\lambda e^{ ik \cdot x}
\eea
one has 
\bea
B= \int \frac{d^3 k }{ (2\pi)^3} \sum_\lambda \left(\varepsilon^\lambda_k  \lambda k A_k^\lambda e^{ik \cdot x} \right)
\eea
and 
\bea
h&=& \int \sum_\lambda  \lambda k \left| A^\lambda_k \right|^2 {d^3 k \over (2\pi)^3} \nonumber \\
&=& \int \frac{ k^3 dk }{2\pi^2 } \left( \left|A^+_k\right|^2 -\left|A^-_k\right|^2 \right) \nonumber \\
&\equiv& \int h_k dk  \; , 
\eea
where we have used the following relations of  the transverse polarization vector: 
$k \cdot \varepsilon_k^\lambda =0$ , $\varepsilon_k^\lambda \cdot (\varepsilon_k^{\lambda^\prime})^* =\delta^{\lambda \lambda^\prime}$, $k \times \varepsilon_k^\lambda = - i \lambda k \varepsilon_k^\lambda$ and $(\varepsilon_k^\lambda)^* =\varepsilon^\lambda_{-k}$.

Similarly one can define the magnetic field energy density as 
\bea
\rho_B &=& \lim_{V\to \infty} \int \frac{d^3 x}{V} \frac{1}{2} B^2  \nonumber \\
&=& \int  \frac { k^4 dk} {4\pi^2}  \left( \left|A^+_k\right|^2 + \left|A^-_k\right|^2 \right) \nonumber \\
&\equiv& \int \rho_k dk  \; .
\eea

Starting from Maxwell equations and the generalized Ohm's law, which reads 
\bea
\nabla \times E + \frac{\partial B} {\partial t} =0 \\
\nabla \cdot E =\rho \\
\nabla \cdot B =0 \\
\nabla \times B - \frac{\partial E}{\partial t} =J  \\
J = \sigma (E + v\times B) + {2\alpha \over \pi } \mu_{X, 5}^{} B 
\eea
one can derive the rate of change for the helicity as
\bea
\frac{\partial h}{\partial t} &=& - \lim_{V\to \infty} \frac{2}{V} \int d^3 x \mathbf{E} \cdot \mathbf{B} \nonumber \\
&=& \lim_{V\to \infty} \int \frac{d^3x}{V} \left( \frac{2}{\sigma} B \cdot \nabla^2 A + \frac{4\alpha}{\pi} \frac{\mu_{X,5}}{\sigma} B^2 \right)  \nonumber \\
&=&- \frac{2}{\sigma} \int k^2 h_k dk + \frac{8 \alpha }{\pi } \frac{\mu_{X, 5}}{\sigma} \int \rho_k d k
\eea
where we have dropped the surface term and wok in the MHD limit.  

Similarly, one can derive the equation of motion for the energy density of the magnetic field 
\bea
\frac{\partial \rho_B^{} } {\partial t} = -\frac{1}{\sigma} \int 2k^2 \rho_k dk + \frac{2\alpha}{\pi} \frac{\mu_{X,5}}{\sigma} \int k^2 h_k d k  \; .
\eea

One can expand results obtained above to the expanding universe by replacing $t$ with the conformal time $\eta$ and replacing the physical quantities $f$ with comoving quantities $f^c$.  Then the equation of motion for $h_k^c$ and $\rho_k^c $ can be written as
\bea
\frac{\partial h_k^c}{\partial \eta } &=& -\frac{2k^2}{\sigma^c } h^c_k + \frac{8\alpha }{\pi } \frac{\mu^c_{X, 5}}{\sigma^c } \rho_k^c  \\
\frac{\partial \rho_k^c}{\partial \eta } &=& -\frac{2k^2}{\sigma^c } \rho^c_k + \frac{2\alpha }{\pi } \frac{\mu^c_{X, 5}}{\sigma^c } k^2 h_k^c  
\eea
where the first term on the right-handed side leads to the decay of the magnetic field and the second term leads to the instability of the chiral plasma. 

\section{Appendix C}
In this section we present transport equations for various SM particles.  Generic equations take the form 
\bea
\dot n_i + 3 H n_i = -S_Y^{} - S^{}_{\rm WS} - S^{}_{SS} + S_Y^{\rm bkg} + S_X^{\rm bkg}+ \cdots
\eea
where the first term on the right-handed side is induced by the Yukawa interaction of the flavor $i$, the second and the third terms are know as the electroweak and strong sphalerons, the term $S^{\rm bkg}$ is induced from the background inputs induced by the hypermagnetic field and other new magnetic field. 

To derive the expression of $S^{\rm bkg}$, we starts from the  axial anomaly of a U(1) gauge field in an expanding universe, 
\bea
\partial_\mu (\sqrt{-g} j^\mu ) = C_X^{}  \frac{\alpha^\prime}{4 \pi} X_{\mu \nu} \widetilde{X}^{\mu\nu } + \cdots
\eea
where $\sqrt{-g} =a^3$ wth $g= {\rm det} (g_{\mu \nu})$,  $\widetilde{X}^{\mu\nu}= 1/2 \varepsilon^{\mu \nu \rho \sigma } X_{\rho \sigma}^{}$ with $X_{\alpha \beta}$ the field strength, $\alpha^\prime$ is the fine structure constants of the new U(1) gauge interactions, $C_X$  incorporates sign and charges of the specific chiral fermion.  Defining the electric and magnetic fields as 
\bea
F_{0i}^{} = aE, \; , F_{ij}^{} = a^2 \varepsilon_{ijk}^{} B^k 
\eea
one has 
\bea
\dot n + 3 H n = C_X^{} {2 \alpha^\prime \over \pi } E \cdot B+ \cdots 
\eea
Introducing comoving electric and magnetic fields as 
\bea
\tilde E = a^2 E \; , \hspace{0.5 cm} \tilde B = a^2 B \; , 
\eea
one has 
\bea
\frac{\partial \tilde n} {\partial \eta } =  C_X^{} \frac{2 \alpha^\prime }{\pi} \tilde E\cdot \tilde B + \cdots
\eea
 where $\tilde n = a^3 n $ and $dt = a d\eta$. 
 
In addition to Eqs (6) and (7), we list in the following transport equations used in our analysis
\bea
-\frac{d}{d \ln T} \left( \frac{\mu_{Q_i} } {T} \right) &=& + \frac{1}{g_{Q_i}} \frac{\gamma_{U_i}} {H} \left(  \frac{\mu_{U_i}}{T} -\frac{\mu_{Q_i} }{T} - \frac{\mu_H}{T}\right)  \\
&&+\frac{1}{g_{Q}} \frac{\gamma_{D_i}} {H} \left(  \frac{\mu_{D_i}}{T} -\frac{\mu_{Q_i} }{T} + \frac{\mu_H}{T}\right)  \nonumber \\
&&-\frac{1}{g_{Q}} \frac{ \gamma_{WS}}{H} \left[ \sum_i {\mu_{L_i} \over T} + 3 \sum_i {\mu_{Q_i} \over T} \right] \nonumber \\
&&-\frac{1}{g_Q}\frac{\gamma_{SS}}{H} \sum_i \left[ 2 \frac{\mu_{Q_i } }{T} - \frac{\mu_{U_i}}{T} - \frac{\mu_{D_i}}{T} \right]  \nonumber \\
-\frac{d}{d \ln T} \left( \frac{\mu_{U_i} } {T} \right) &=& - \frac{1}{g_{U}} \frac{\gamma_{U_i}} {H} \left(  \frac{\mu_{U_i}}{T} -\frac{\mu_{Q_i} }{T} - \frac{\mu_H}{T}\right) \\
&&+\frac{1}{g_U} \frac{\gamma_{SS}}{H}\sum_i \left[ 2 \frac{\mu_{Q_i } }{T} - \frac{\mu_{U_i}}{T} - \frac{\mu_{D_i}}{T} \right]   \nonumber\\
-\frac{d}{d \ln T} \left( \frac{\mu_{D_i} } {T} \right) &=& - \frac{1}{g_{D}} \frac{\gamma_{D_i}} {H} \left(  \frac{\mu_{D_i}}{T} -\frac{\mu_{Q_i} }{T} + \frac{\mu_H}{T}\right) \\
&&+\frac{1}{g_D} \frac{\gamma_{SS}}{H}\sum_i \left[ 2 \frac{\mu_{Q_i } }{T} - \frac{\mu_{U_i}}{T} - \frac{\mu_{D_i}}{T} \right]  \nonumber\\
-\frac{d}{d \ln T} \left( \frac{\mu_{H} } {T} \right) &=& \frac{1}{g_{H}}  \sum_i  \left[ \frac{\gamma_{U_i}} {H} \left(  \frac{\mu_{U_i}}{T} -\frac{\mu_{Q_i} }{T} - \frac{\mu_H}{T}\right)  \right. \nonumber \\
&&- \frac{\gamma_{D_i}} {H} \left(  \frac{\mu_{D_i}}{T} -\frac{\mu_{Q_i} }{T} + \frac{\mu_H}{T}\right)  \nonumber \\
&&- \left. \frac{\gamma_{E_i}} {H} \left(  \frac{\mu_{E_i}}{T} -\frac{\mu_{L_i} }{T} + \frac{\mu_H}{T}\right)  \right]
\eea
where $g_X$ is the degree of freedom for species ``i". $\gamma_{U_i}$ and $\gamma_{D_i}$ are Yukawa interaction rate for up-type and down-type quarks respectively, $\gamma_{WS}$ and $\gamma_{SS}$ are rates for electroweak sphaleron and strong sphaleron processes, separately.

\bibliography{references}

\end{document}